\documentclass[a4paper,fleqn,10pt]{article}
\pdfoutput=1
\usepackage{amsmath}
\usepackage{amssymb}
\usepackage{array}
\usepackage{calc}
\usepackage{longtable}
\usepackage{multirow}
\usepackage{lineno}
\usepackage{pstricks}
\usepackage{graphicx}
\usepackage{xspace}
\usepackage{units}

\numberwithin{equation}{section}
\usepackage{cite}
\usepackage[pdfborder={0 0 0}]{hyperref}
\usepackage[format=hang,labelfont=bf,hypcap=true]{caption}
\usepackage{subcaption}
\usepackage{rotating}
\usepackage{units}
\usepackage{sectsty}
\allsectionsfont{\sffamily}
\subsubsectionfont{\mdseries\itshape\large}
\makeatletter
\DeclareRobustCommand*{\bfseries}{%
  \not@math@alphabet\bfseries\mathbf
  \fontseries\bfdefault\selectfont
  \boldmath
}
\makeatother
\let\spreprint\empty
\newcommand{\preprint}[1]{\def\spreprint{\protect#1}}
\let\sinstitute\empty
\newcommand{\institute}[1]{\def\sinstitute{\protect#1}}
\makeatletter
\renewcommand{\maketitle}{\begingroup
  \null\thispagestyle{empty}%
    \ifx\spreprint\empty
      \vskip 5ex
    \else
      \flushright\large\spreprint\vskip 2ex
    \fi
    \vskip 5ex
    \flushleft
      {\sffamily\bfseries\huge\@title}\vskip 6ex
      \@author\vskip 2ex
      \ifx\sinstitute\empty
      \else
        {\small\sinstitute}
      \fi
    \vskip 5ex
  \endgroup
}
\makeatother
\renewenvironment{abstract}{\begin{center}
  {\large\sffamily\bfseries Abstract: }
  \begin{minipage}[t]{0.75\textwidth}
}{\end{minipage}\end{center}\vskip 10ex}

\numberwithin{equation}{section}

\newcommand{\POWHEG}{P\protect\scalebox{0.8}{OWHEG}\xspace}

\newcommand{\SMCatNLO}{S--M\protect\scalebox{0.8}{C}@N\protect\scalebox{0.8}{LO}\xspace}
\newcommand{\MCatNLO}{M\protect\scalebox{0.8}{C}@N\protect\scalebox{0.8}{LO}\xspace}

\newcommand{\MENLOPS}{ME\protect\scalebox{0.8}{NLO}PS\xspace}
\newcommand{\MEPSatNLO}{M\protect\scalebox{0.8}{E}P\protect\scalebox{0.8}{S}@N\protect\scalebox{0.8}{LO}\xspace}
\newcommand{\CKKW}{CKKW\xspace}

\newcommand{\OpenLoops}{O\protect\scalebox{0.8}{PEN}L\protect\scalebox{0.8}{OOPS}\xspace}
\newcommand{\SherpaOpenLoops}{S\protect\scalebox{0.8}{HERPA}+O\protect\scalebox{0.8}{PEN}L\protect\scalebox{0.8}{OOPS}\xspace}
\newcommand{\Collier}{C\protect\scalebox{0.8}{OLLIER}\xspace}
\newcommand{\VBFNLO}{V\protect\scalebox{0.8}{BF}N\protect\scalebox{0.8}{LO}\xspace}
\newcommand{\Sherpa}{S\protect\scalebox{0.8}{HERPA}\xspace}
\newcommand{\Comix}{C\protect\scalebox{0.8}{OMIX}\xspace}

\newcommand{\Amegic}{A\protect\scalebox{0.8}{MEGIC++}\xspace}

\newcommand{\ATLAS}{ATLAS\xspace}
\newcommand{\CMS}{CMS\xspace}
\newcommand{\SecRef}[1]{Section~\ref{#1}}

\long\def\symbolfootnote[#1]#2{\begingroup%
\def\thefootnote{\fnsymbol{footnote}}\footnote[#1]{#2}\endgroup}

\newcommand{\sbr}[1]{\left[ #1\right]}

\newcommand{\done}{{\rm d}}
\newcommand{\order}{\mathcal{O}}
\newcommand{\mc}[1]{\mathcal{#1}}
\newcommand{\mr}[1]{\mathrm{#1}}

\newcommand{\bea}{\begin{eqnarray}}
\newcommand{\eea}{\end{eqnarray}}
\newcommand{\bi}{\begin{itemize}}
\newcommand{\ei}{\end{itemize}}

\hypersetup{
  pdfauthor={S. Hoeche, F. Krauss, S. Pozzorini, M. Schoenherr, J. M. Thompson, K. C. Zapp},
  pdftitle={Triple vector boson production through Higgs-Strahlung with NLO multijet merging},
  pdfkeywords={QCD, NLO, Higgs}
}
\preprint{
  SLAC-PUB 15933\\ IPPP/14/23\\ DCPT/14/46\\ 
  LPN14-060\\ ZU-TH 13/14\\ CERN-PH-TH/2014-051\\ MCNET-14-007\\ 
  HT 14-008}
\author{
  S.~H\"oche$^1$, F.~Krauss$^2$, S.~Pozzorini$^3$, 
  M.~Sch\"onherr$^2$, J.~M.~Thompson$^2$, K.~C.~Zapp$^4$
}
\title{Triple vector boson production through Higgs-Strahlung
  with NLO multijet merging}
\institute{
  $^1$ SLAC National Accelerator Laboratory, Menlo Park, CA 94025, USA\\
  $^2$ Institute for Particle Physics Phenomenology,
  Durham University, Durham DH1 3LE, UK\\
  $^3$ Physik--Institut, Universit{\"a}t Z{\"u}rich, CH--8057 Z{\"u}rich, Switzerland\\
  $^4$ CERN, Department of Physics, CH--1211 Geneva 23, Switzerland\\
}
\begin{document}
\maketitle
\begin{abstract}
  Triple gauge boson hadroproduction, in particular the production of three
  $W$-bosons at the LHC, is considered at next-to leading order accuracy in 
  QCD.  The NLO matrix elements are combined with parton showers.  Multijet 
  merging is invoked such that NLO matrix elements with one additional jet
  are also included.  The studies here incorporate both the signal and all 
  relevant backgrounds for $VH$ production with the subsequent decay of the 
  Higgs boson into $W$-- or $\tau$--\-pairs. They have been performed  
  using \SherpaOpenLoops in combination with \Collier.
\end{abstract}

\section{Introduction}
\label{Sec:Introduction}
The imminent second round of data taking at the LHC presents new opportunities
for studying physics at the electroweak to TeV scale.  In light of 
the recent discovery of a Higgs boson~\cite{Aad:2012tfa,Chatrchyan:2012ufa}, 
with all experimental determinations of its properties up to now compatible 
with Standard Model expectations based on the Brout--\-Englert--\-Higgs (BEH)
mechanism~\cite{Englert:1964et,Higgs:1964ia,Higgs:1964pj,Guralnik:1964eu},
it is clear that increasingly precise studies become necessary in order to
look for subtle effects where new physics could manifest itself.

A prime candidate for such studies is the production of multiple gauge bosons:
channels involving $ZZ$, $WW$ and $\gamma\gamma$ final states have been 
employed, among others, for the discovery of the Higgs boson, while processes 
with $W\gamma$, $WZ$, $ZZ$, and $Z\gamma$ final states are frequently used by 
the experiments to search for anomalous triple gauge boson couplings, see
for instance~\cite{Aad:2011xj,Aad:2011cx,Aad:2012mr,Chatrchyan:2013fya}.  
Clearly, with higher energies, such searches can and will be extended to also 
include anomalous quartic gauge couplings.  In addition, multi-boson channels, 
and in particular those that lead to final states involving three leptons, are 
important backgrounds in searches for new particles; as illustrative example 
consider neutralino-chargino pair production and their subsequent decay in 
supersymmetric extensions of the Standard Model. 

This publication focuses on the production of a Standard Model Higgs 
boson in the Higgsstrahlung process (associated $VH$ production) and its 
subsequent decay into $W$-- or $\tau$--\-pairs.  Apart from the signal, all 
relevant background channels will be studied as well. This includes multiple 
gauge bosons final states such as $WZ$, $WWW$, $ZWW$, $ZZ$, $WZZ$ and $ZZZ$.  
The studies presented here follow closely the recent analyses by \ATLAS and 
\CMS~\cite{TheATLAScollaboration:2013hia,ATLAS:2012toa,CMS:zwa}.

In many of these processes, QCD corrections play a significant role, from highly
phase-space dependent $K$-factors ranging between 1.5 and 2 to the fact that
the emergence of additional jets can be used to shed light on the actual 
production mechanism giving rise to triple gauge boson final states.  In 
addition, quite often vetoing additional jets is a very good way to suppress
unwanted backgrounds, a prime example being the massive suppression of the 
$t_{\to W^+b}\bar t_{\to W^-\bar b}W$ background to $WWW$ production or other signals,
which allows us here to ignore this class of processes. 

For the signal process, $VH$--associated production, parton--\-level results
are available at next-to leading order accuracy (NLO) in the perturbative
expansion of QCD~\cite{Han:1991ia} and NNLO results are known for
more than a decade~\cite{Brein:2003wg,Ferrera:2011bk}.
Resummed predictions were computed more recently~\cite{Dawson:2012gs}.
The NLO QCD corrections to triple gauge boson production have first been 
calculated in~\cite{Lazopoulos:2007ix,Binoth:2008kt}, the leptonic decay 
of the bosons has been discussed in~\cite{Hankele:2007sb,Campanario:2008yg} 
and it has also been implemented in the \VBFNLO code~\cite{Arnold:2011wj}.
Predictions at NLO QCD for triple gauge boson production in association with 
one extra jet are presented for the first time in this paper.

For the calculation of the virtual corrections 
we employ \OpenLoops~\cite{Cascioli:2011va}, a fully automated one-loop generator based on 
a fast numerical recursion for multi-particle processes.
For tensor and scalar integrals we use
the \Collier library~\cite{collier}, which guarantees high numerical stability thanks to the
methods of~\cite{Denner:2002ii,Denner:2005nn,Denner:2010tr}.
For the Born and real emission contributions the matrix element generators
\Amegic~\cite{Krauss:2001iv} and \Comix~\cite{Gleisberg:2008fv} are used.  The 
mutual cancellation of infrared divergences in real and virtual contributions 
is achieved through the dipole formalism~\cite{Catani:1996vz,Catani:2002hc}
and its automated implementation in both \Amegic~\cite{Gleisberg:2007md}
and \Comix. The overall event generation is handled by 
\Sherpa~\cite{Gleisberg:2003xi,Gleisberg:2008ta}. For the first 
time, the NLO QCD calculations are combined consistently with parton 
showers, employing the \SMCatNLO variant~\cite{Hoeche:2011fd,Hoeche:2012ft} of 
\MCatNLO~\cite{Frixione:2002ik,Frixione:2003ei}.  Parton showers are generated 
by \Sherpa, based on Catani--\-Seymour dipole subtraction~\cite{Catani:1996vz,Catani:2002hc}
as suggested in~\cite{Nagy:2005aa} and implemented in~\cite{Schumann:2007mg}.
This setup for the matching was recently employed $W\!+3$ jets production 
\cite{Hoeche:2012ft}, dijet production \cite{Hoeche:2012fm}, and for $t\bar{t}b\bar{b}$
production in~\cite{Cascioli:2013era}.  In addition, a multijet merging with 
NLO matrix elements including one additional jet is included, following the 
\MEPSatNLO algorithm~\cite{Gehrmann:2012yg,Hoeche:2012yf}.  This method has 
recently been employed for a number of processes, among them top pair 
production with up to two jets~\cite{Hoeche:2013mua,Hoeche:2014qda}, 
Higgs production in gluon fusion with up to two jets~\cite{Hoeche:2014lxa} and, 
similarly, the production of $4$ leptons in association with up to one
jet~\cite{Cascioli:2013gfa}\footnote{
  There are, of course, other matching algorithms such as the \POWHEG method
  described in~\cite{Nason:2004rx,Frixione:2007vw}, and also merging 
  algorithms, both for matrix elements at leading 
  order~\cite{Catani:2001cc,Lonnblad:2001iq,Krauss:2002up,Mangano:2001xp,
    Alwall:2007fs,Hamilton:2009ne,Hoeche:2009rj}
  and at next-to leading order~\cite{Lonnblad:2012ix,Frederix:2012ps,
    Platzer:2012bs}.
}.

The Monte-Carlo methods used to simulate jet production and evolution 
are discussed in Sec.\ \ref{sec:framework}.  
Section \ref{sec:results} presents results obtained with the \SMCatNLO 
matching and \MEPSatNLO merging methods.  We focus on the treatment 
of signal and background with typical cuts as used by \ATLAS and \CMS,
\cite{TheATLAScollaboration:2013hia,ATLAS:2012toa,CMS:zwa}.
This publication closes with a summary and some outlook in 
\SecRef{Sec:Conclusions}.

\section{Matching and merging techniques in 
  \texorpdfstring{\protect\Sherpa}{Sherpa}}
\label{sec:framework}
This section reviews the basic MC event generation techniques used
in our analysis. We focus on new developments in matching and 
merging methods, which allow to combine fixed-order NLO
calculations and parton shower simulations.

\subsection{\texorpdfstring{\protect\SMCatNLO}{S-MC@NLO}}
\label{sec:smcatnlo}

Leading order cross sections, including the subsequent parton shower
evolution in the initial and final state can schematically be written as
\begin{equation}\label{eq:match_lo}
  \done\sigma^{\rm(PS)}\,=\,
  \done\Phi_B\,\mr{B}_n(\Phi_B)\,\mc{F}_{n}(\mu_Q^2)\;,
\end{equation}
where $\done\Phi_B$ denotes the phase space element for the 
Born--\-level kinematics and $\mr{B}_n(\Phi_B)$ is the Born--\-level 
differential cross section with $n$ external partons, composed of the 
corresponding parton--\-level cross section at leading order, convoluted 
with the PDFs and multiplied with suitable symmetry and flux factors.  The 
parton shower evolution of the $n$--\-parton configuration is encoded in the 
generating functional, $\mc{F}_n(\mu_Q^2)$ with the resummation scale 
$\mu_Q$. This scale is a free parameter, entering in addition to the 
renormalization and factorization scales $\mu_R$ and $\mu_F$, and it may be 
chosen in a process--\-dependent way. At leading order, $\mu_Q=\mu_F$.
The parton--\-shower generating functional is defined by splitting kernels 
$\mr{K}_n$, and the corresponding Sudakov form factor 
$\Delta_n(t^\prime,t)\,=\,\exp\big[-\int_{t'}^{\,t}\,\done\Phi_1\,\mr{K}_n(\Phi_1)\big]$
\begin{equation}\label{eq:gen_ps}
  \mc{F}_n(t)\,=\,
  \Delta_n(t_c,t)\,+\,
  \int_{t_c}^t\done\Phi_1^\prime\,
  \mr{K}_n(\Phi_1^\prime)\,\Delta_n(t^\prime,t)\,\mc{F}_{n+1}(t')\,,
\end{equation}
The emission phase space is parametrized in terms of evolution parameter $t$,
splitting variable $z$, and azimuthal angle $\phi$ as $\done\Phi_1\,=\,
\done t\,\done z\,\done\phi\,J(t,z,\phi)$,
where $J(t,z,\phi)$ denotes a Jacobian factor. $t_c$ is the infrared cut-off
of the parton shower, typically of the order of a GeV.  Equation~\eqref{eq:gen_ps} 
simultaneously describes the probability of no further parton emission (first 
term) and a single emission at scale $t^\prime$ (second term). If 
such an emission takes place, Eq.~\eqref{eq:gen_ps} is iterated, with the 
boundary conditions set by the newly formed partonic state.

By now, there are two different classes of algorithms to promote the leading 
order (LO) expression in Eq.~\eqref{eq:match_lo} to NLO accuracy, namely the
\POWHEG method~\cite{Nason:2004rx,Frixione:2007vw} and the \MCatNLO matching 
method~\cite{Frixione:2002ik}.  In \MCatNLO, the parton-shower approximation
is used to obtain universal subtraction terms that can be used to cancel the
singularities in real and virtual corrections. This method was extended 
in~\cite{Hoeche:2011fd,Hoeche:2012ft,Hoeche:2012fm} by modifying the parton 
shower such that its splitting kernels for the first emission include the full 
color and spin dependence present in the real corrections. This may introduce
color--\-suppressed but logarithmically enhanced contributions into the Sudakov 
form factor. This method is referred to as \SMCatNLO in the following.
Its cross section is computed as 
\begin{equation}\label{eq:mcatnlo}
  \done\sigma^\text{(\SMCatNLO)}\,=\;
  \done\Phi_B\,\bar{\mr{B}}_n(\Phi_B)\,\bar{\mc{F}}_n(\mu_Q^2)\,+\,
  \done\Phi_R\,\mr{H}_{n}(\Phi_R)\,\mc{F}_{n+1}(\tilde\mu_Q^2)\;,
\end{equation}
where the combination of $\bar{\mr{B}}$ and $\mr{H}$ provides
the exact next-to leading order cross section and the functionals
$\mc{F}_n$ and $\bar{\mc{F}}_n$ change its phase-space dependence,
but not the normalization. $\mr{H}_n$ captures
the subtracted real-emission contribution, while $\bar{\mr{B}}_n$
contains all other terms, projected onto Born kinematics.
\begin{equation}\label{eq:mcatnlo_bbar_h}
  \begin{split}
  \bar{\mr{B}}_n(\Phi_B)\,=&\;\mr{B}_n(\Phi_B)+\tilde{\mr{V}}_n(\Phi_B)
  +\mr{I}_n(\Phi_B,\,\mu_Q^2)\,,\\
  \mr{H}_{n}(\Phi_R)\,=&\;\mr{R}_{n}(\Phi_R)
  -\mr{D}_{n}(\Phi_R)\,\Theta\left(\mu_Q^2-t\right)\;.
  \end{split}
\end{equation}
In addition to the squared Born matrix element $\mr{B}_n(\Phi_B)$ the virtual 
and real corrections, $\tilde{\mr{V}}_n(\Phi_B)$ and $\mr{R}_{n}(\Phi_R)$ 
respectively, have been introduced, the latter together with the corresponding 
phase space element $\done\Phi_R$.  This phase space element factorizes as 
$\Phi_R\,=\,\Phi_B\times\Phi_1$, which is used to facilitate integration of
the subtraction terms $\mr{D}_{n}(\Phi_B,\Phi_1)$ over the one--\-particle 
emission phase space and define the integrated subtraction terms 
$\mr{I}_n(\Phi_B)$~\cite{Catani:1996vz,Catani:2002hc},
\begin{equation}\label{eq:isub}
  \mr{I}_n(\Phi_B,\,\mu_Q^2)\,=\,
  \int\done\Phi_1\,\mr{D}_{n}(\Phi_B,\Phi_1)\,
  \Theta\left(\mu_Q^2-t\right)\;.
\end{equation}
Equation~\eqref{eq:isub} is known analytically in $d$ dimensions for 
$\mu_Q^2\to\infty$, as is necessary to extract the poles in the dimensional
regularization parameter $\varepsilon$. The value for finite $\mu_Q^2$ is
computed by calculating the finite remainder in $d=4$ dimensions with
Monte-Carlo techniques~\cite{Hoeche:2012ft,Hoeche:2012fm}.

The generating functional of the \SMCatNLO is defined as
\begin{equation}\label{eq:gen_mcatnlo}
  \bar{\mc{F}}_{n}(t)\,=\;
    \bar{\Delta}_n(t_c,t)\,+\,
    \int_{t_c}^{t}\!\done\Phi_1^\prime\,
    \frac{\mr{D}_{n}(\Phi_B,\Phi_1^\prime)}{\mr{B}_n(\Phi_B)}\,
    \bar{\Delta}_n(t^\prime,t)\,\mc{F}_{n+1}(t^\prime)\;.
\end{equation}
It differs from the parton-shower expression, Eq.~\eqref{eq:gen_ps},
by sub-leading color contributions and spin correlation effects.
Note that all secondary emissions are treated by a standard 
parton shower, indicated by $\mc{F}_{n+1}(t')$.

\subsection{\texorpdfstring{\protect\MENLOPS}{MENLOPS}}
\label{sec:menlops}
The method outlined above can be improved with higher-order tree-level 
calculations using a multi-jet merging technique~\cite{Hamilton:2010wh,Hoeche:2010kg,Gehrmann:2012yg}.  
A merging scale, $Q_\text{cut}$, is introduced, like in the
pure LO merging algorithms~\cite{Catani:2001cc,Lonnblad:2001iq,Krauss:2002up,Mangano:2001xp,
    Alwall:2007fs,Hamilton:2009ne,Hoeche:2009rj},
which restricts the phase space of emissions in the parton shower from above,
and emissions in the matrix elements from below.

The restricted \MCatNLO simulation for the ``core'' process with $n$ particles 
generates the following terms
\begin{equation}\label{eq:nlo_term_base}
  \begin{split}
    \done\sigma_{n}^{\rm excl}
    \,=&\;\done\Phi_{n}\;
          \bar{\mr{B}}_{n}(\Phi_n)\,
          \bar{\mc{F}}_{n}(\mu_Q^2\,;<\!Q_{\rm cut})\\
    &{}+\done\Phi_{n+1}\,
         \Theta(Q_{\rm cut}-Q(\Phi_{n+1}))\,
         \mr{H}_{n}(\Phi_{n+1})\,
         \mc{F}_{n+1}(\mu_Q^2\,;<\!Q_{\rm cut})\;,
  \end{split}
\end{equation}
where $\bar{\mc{F}}_{n}(\mu_Q^2\,;<\!Q_{\rm cut})$ is the functional 
of the vetoed \SMCatNLO, and $\mc{F}_{n+1}(\mu_Q^2\,;<\!Q_{\rm cut})$ 
is the functional of the truncated vetoed parton shower~\cite{
  Nason:2004rx,Frixione:2007vw,Hoeche:2009rj}.

The next higher jet multiplicities are calculated at leading order accuracy.
A local $K$-factor is applied to preserve the total cross section to NLO
accuracy,
\begin{equation}
  \begin{split}\label{eq:lo-term}
    \done\sigma_{n+k}
    \,=&\;\done\Phi_{n+k}\,\Theta(Q(\Phi_{n+k})-Q_{\rm cut})\;
          k_n\!\left(\Phi_{n+1}(\Phi_{n+k})\right)\;\mr{B}_{n+k}(\Phi_{n+k})\;
          \mc{F}_{n+k}(\mu_Q^2\,;<\!Q_{\rm cut})\;.
  \end{split}
\end{equation}
Here $\Phi_{n+1}(\Phi_{n+k})$ is defined by the kinematics mapping
of the parton shower, and
\begin{equation}
  \begin{split}\label{eq:loc-k-fac}
  k_n(\Phi_{n+1})\,=\;
	 \frac{\bar{\rm B}_n(\Phi_n)}{{\rm B}_n(\Phi_n)}
	 \left(
	       1-\frac{\mr{H}_n(\Phi_{n+1})}{{\rm R}_n(\Phi_{n+1})}
	 \right)+\frac{\mr{H}_n(\Phi_{n+1})}{{\rm R}_n(\Phi_{n+1})}\;.
  \end{split}
\end{equation}
is the local $K$-factor \cite{Gehrmann:2012yg}. It is constructed such that a sample where exactly 
one jet at leading order accuracy is merged on top of the underlying \SMCatNLO 
reproduces this \SMCatNLO except for potential sub-leading color 
corrections in the \SMCatNLO $n$-jet simulation versus the 
showered $n+1$-jet simulation.

\subsection{\texorpdfstring{\protect\MEPSatNLO}{MEPS@NLO}}
\label{sec:mepsatnlo}

The above merging method can be extended to the next-to-leading order
also for the $n+k$-jet exclusive simulations. In order not to spoil
the NLO-accuracy of these simulations it is not enough to simply implement
a truncated vetoed parton shower as this is done in leading-order merging.
The first-order expansion of the vetoed shower would generate corrections
of order $\alpha_s$, which must be subtracted. This leads to the following 
expression for the differential cross section in the $n+k$-jet sample:
\begin{equation}\label{eq:nlo_term}
  \begin{split}
    \done\sigma_{n+k}^{\rm excl}
    \,=&\;\done\Phi_{n+k}\,\Theta(Q(\Phi_{n+k})-Q_{\rm cut})\;
          \tilde{\mr{B}}_{n+k}(\Phi_n+k)\,
          \bar{\mc{F}}_{n+k}(\mu_Q^2\,;<\!Q_{\rm cut})\\
    &+\done\Phi_{n+k+1}\,\Theta(Q(\Phi_{n+k})-Q_{\rm cut})\,
         \Theta(Q_{\rm cut}-Q(\Phi_{n+k+1}))\,
         \tilde{\mr{H}}_{n+k}(\Phi_{n+k+1})\,
         \mc{F}_{n+k+1}(\mu_Q^2\,;<\!Q_{\rm cut})\;,
  \end{split}
\end{equation}
The extended subtraction is implemented by the modified differential
cross sections $\tilde{\mr{B}}_{n+k}(\Phi_{n+k})$ and 
$\tilde{\mr{H}}_{n+k}(\Phi_{n+k+1})$, defined as
\begin{equation}\label{eq:mcatnlon_bbar_h}
  \begin{split}
    \tilde{\rm B}_{i}(\Phi_{i})=&\;
      {\rm B}_{i}(\Phi_{i})+\tilde{\mr{V}}_{i}(\Phi_{i})
        +{\rm I}_{i}(\Phi_{i})
      +\int\done\Phi_1\,\sbr{\tilde{\rm D}_{i}(\Phi_{i},\Phi_1)
        -{\rm D}_{i}(\Phi_i,\Phi_1)}\\
    \tilde{\mr{H}}_{i}(\Phi_{i+1})=&\;
    \mr{R}_{i}(\Phi_{i+1})-
      \tilde{\mr{D}}_{i}(\Phi_{i+1})\;,\\
  \end{split}
\end{equation}
which take the probability of truncated parton shower emissions into
account~\cite{Hoeche:2012yf,Gehrmann:2012yg}. To this end, the dipole
terms used in the \SMCatNLO are extended by the parton-shower emission
probabilities, $\mr{B}_{i}(\Phi_i)\,\mr{K}_{j}(\Phi_{1,i+1})$, where
$\mr{K}_{j}(\Phi_{1,i+1})$ is the sum of all shower splitting functions
for the intermediate state with $j<i$ in a predefined shower tree
which leads to the final state with kinematical configuration $\Phi_i$. Thus, 
\begin{equation}\label{eq:compound_kernel}
   \tilde{\mr{D}}_{i}(\Phi_{i+1})\,=\;
   \mr{D}_{i}(\Phi_{i+1})\,\Theta(t_{i}-t_{i+1})\;
   +\sum_{j=0}^{i-1}\mr{B}_{i}(\Phi_i)\,\mr{K}_{j}(\Phi_{1,i+1})\,
      \Theta(t_j-t_{i+1})\,\Theta(t_{i+1}-t_{j+1})\,\Big|_{\,t_0=\mu_Q^2}\;.
\end{equation}
This expression has a simple physical interpretation: The first term 
corresponds to the coherent emission of a parton from the external $i$-parton 
final state.  It contains all soft and collinear singularities which are
present in the real-emission matrix elements.  The sum in the second term
corresponds to emissions from the intermediate states with $i$ partons and
in fact the terms $\mr{B}_{i}(\Phi_i)\,\mr{K}_{j}(\Phi_{1,i+1})$ stem from the
expansion of the Sudakov form factor of the truncated shower to first order
in the strong coupling.  All these terms can be implemented in the parton 
shower approximation, because soft divergences are regulated by the finite 
mass of the intermediate particles.  

\section{Results}
\label{sec:results}

\subsection{Details of the analyses}
\label{sec:res-ana}

There are current efforts from both \CMS and \ATLAS to search for the trilepton ($\ell=e,\mu$)
final states emerging from $WH$--\-associated production, where the Higgs boson 
decays either into $\tau$ or $W$ pairs~\cite{TheATLAScollaboration:2013hia,
  ATLAS:2012toa,CMS:zwa}\footnote{
  Note that the \ATLAS publication also includes similar searches in 
  $ZH$--associated production which will not be considered here.
}. 
These final states allow a direct probe of the coupling between the Higgs 
boson and the weak bosons. In the following we present two analyses: the 
first inspired by a recent search by the \CMS collaboration~\cite{CMS:zwa}, 
the second following searches from the \ATLAS 
collaboration~\cite{TheATLAScollaboration:2013hia,ATLAS:2012toa}.
The majority of the cuts that
are applied in both are given in Tab.\ \ref{table:analysis_cuts}. 
Their crucial features in reducing unwanted backgrounds are a veto on $Z$ 
bosons, which is realized differently in both analysis, and vetoes on jet 
activity to eliminate the large background from $t\bar{t}V$ production.  
Jets are reconstructed in both analyses using the anti-$k_T$ 
algorithm~\cite{Cacciari:2008gp,Cacciari:2011ma} with the parameters given 
in Tab.\ \ref{table:analysis_cuts}. In the \ATLAS-inspired 
analysis, events are allowed to contain at most one jet, which must not be 
a $b$-jet. The \CMS-inspired analysis vetoes all events with a jet of 
$p_\perp>\unit[40]{GeV}$ and any containing $b$-jets. Both analyses dress 
electrons with all surrounding photons within a cone of $\Delta R=0.1$ 
while muons are left bare.

The \ATLAS-inspired analysis requires exactly three isolated leptons of 
net charge $\pm 1$. At least one of the leptons needs to have a transverse 
momentum of more than \unit[25]{GeV} for electrons and \unit[21]{GeV} for 
muons, the other two leptons $p_\perp>\unit[10]{GeV}$ each. They are 
labeled in the following way: the lepton with charge different from the 
others is called lepton~0, of the two others the one with smaller distance 
$\Delta R$ from lepton~0 is called lepton~1 and the remaining one is labeled as 
lepton~2. The leptons are considered isolated if the transverse energy of all 
visible particles in a cone of radius $\Delta R_\text{iso} = 0.2$ for 
leptons~0 and 1 and $\Delta R_\text{iso} = 0.4$ for lepton~2 around the lepton 
is less than \unit[10]{\%} of the lepton $p_\perp$.  
After this pre-selection events containing a same-flavor-opposite-sign (SFOS) 
lepton pair are classified as $Z$ enriched, those that do not belong to the $Z$ 
depleted sample.  In this publication only the $Z$~depleted subsample is 
considered. Contrary to the experimental analysis in \cite{ATLAS:2012toa} 
no requirement on the missing transverse energy is applied.

The \CMS-inspired analysis on the contrary requires at least three isolated 
leptons of net charge $\pm 1$. Of those, at least one is required to have 
$p_\perp>\unit[20]{GeV}$ while the others must only fulfil 
$p_\perp>\unit[10]{GeV}$. The lepton isolation in turn depends on lepton 
flavor rather than classification. Electrons are considered isolated if 
in a cone of radius $\Delta R_\text{iso}=0.4$ the sum of the transverse 
energy of all visible particles does not exceed \unit[15]{\%} of the lepton 
$p_\perp$, while muons must satisfy this limit only in a cone of size 
$\Delta R_\text{iso}=0.3$. In case a pair of same-flavor-opposite-sign (SFOS) 
leptons is present in the event, the event is discarded if its invariant mass 
is closer to the nominal $Z$ boson mass than \unit[25]{GeV}.

\begin{table}
  \begin{center}
    \begin{tabular}{|l|c|c|}
      \hline
      Cut & \ATLAS & \CMS $\vphantom{\frac{1^|}{2^|}}$\\
      \hline
      $p_{\perp,\,{\mathrm{min}}}^\ell$ 
        & \qquad\qquad \unit[10]{GeV} \qquad\qquad{} & \qquad\qquad \unit[10]{GeV} \qquad\qquad{} \\
      $|\eta^e_{\mathrm{max}}|$ 
        & 2.47 & 2.5 \\
      $|\eta^\mu_{\mathrm{max}}|$ 
        & 2.5 & 2.4 \\
      $N_\mathrm{leptons}$ 
        & 3 & $\ge$3 \\
      $Z$ veto 
        & no SFOS & $|m_Z-m_\mathrm{SFOS}|>\unit[25]{GeV}$\\
      $|\sum Q_\ell|$\quad 
        & $+1$ & $+1$ \\
      Jet $p_{\perp,\mathrm{min}}$ 
        & \unit[25]{GeV} & \unit[20]{GeV} \\
      Jet d$R$ 
        & 0.4 & 0.5 \\   
      $E_{\perp,\,\mathrm{min}}^\mathrm{miss}$ 
        & -- &  \unit[40]{GeV} \\
      \hline
    \end{tabular}
    \caption{
	      Cuts for the \protect\ATLAS- and \protect\CMS-inspired analyses.
	      \label{table:analysis_cuts}
	    }
  \end{center}
\end{table}

Both the \ATLAS and \CMS analyses include regions with more cuts than are 
described here, however the observables presented do not use these regions.

\subsection{Monte Carlo samples}
\label{sec:res-mc}

We consider $pp\to 3\ell+E_\perp^\text{miss}+X$ production at the 
LHC at a center-of-mass energy of \unit[8]{TeV}.
All processes with at least three leptons that involve an on--\-shell 
Higgs boson are considered as signal processes, and those 
which do not are considered background processes. Neutrinos do not 
necessarily need to be present as missing transverse energy can also 
be generated due to the limited detector acceptance in rapidity. 

The signal is comprised primarily of  $W^\pm H(\rightarrow W^+W^-)$, 
$W^\pm H(\rightarrow \tau^+\tau^-)$ and $ZH(\rightarrow W^+ W^-)$, but 
includes also $ZH(\rightarrow \tau^+\tau^-)$, $W^\pm H(\rightarrow ZZ)$ 
and $ZH(\rightarrow ZZ)$ as subdominant contributions. All signal 
processes are calculated at \MEPSatNLO accuracy, merging the respective 
processes accompanied by zero/one jets at NLO and by two jets at LO 
accuracy.
The background processes considered are high multiplicity bosonic final 
states: $W^\pm Z$, $ZZ$, $W^\pm W^+W^-$, $W^+W^-Z$, $W^\pm ZZ$ and $ZZZ$, which can evade the $Z$ veto by the same method as $W^\pm Z$, and also include hadronic decays of the bosons.
Higher multiplicity final states do not have a significant enough 
contribution to be considered.  In addition, the production of an 
off--\-shell Higgs boson decaying to an on--\-shell $V$ boson pair is also 
considered as part of the background. The cross section for this process 
is very small as compared to the production of the on--\-shell Higgs boson, 
and it contributes mostly through its interference with the triple boson 
background. The $W^\pm Z$ boson background remains dominant over large portions 
of phase space; this is due to lost leptons and, more importantly, due to 
decays into $\tau$-leptons which enable the evasion of the $Z$ veto. Of less 
importance is the $W^\pm W^+W^-$ process, nonetheless warranting high 
theoretical accuracy. Thus, both $W^\pm Z$ and $W^\pm W^+W^-$ are calculated at 
the same accuracy as the signal processes, while the remaining subdominant 
background processes, $ZZ$, $W^+W^-Z$, $W^\pm ZZ$ and $ZZZ$,
are considered at \MENLOPS accuracy, i.e.\ NLO accuracy for the respective 
inclusive process and leading order accuracy when the gauge bosons are 
accompanied by one and two jets. Further, in order to prevent 
$tV_1V_2$/$\bar{t}V_1V_2$ contributions entering the $V_1V_2 W^\pm j$ 
calculation, and  $t\bar{t}V$ contributions entering the $V W^+W^-jj$ 
calculation, only light quarks are considered in the matrix element final 
state. 

The Higgs and $W/Z$ gauge boson decays are treated in the narrow width 
approximation, including spin correlation effects throughout all decay 
chains. The kinematics are then corrected by redistributing the boson's
propagator mass onto a Breit-Wigner distribution. In cases where $1\to 2$ 
decays are not allowed kinematically their $1\to 3$ substructure is resolved. 
This is relevant mainly for $H\to VV^*$ decays.
Additionally, all decays receive higher-order QCD and QED corrections 
through intermediate parton showering or YFS-type soft-photon resummation 
(including full $\order(\alpha)$ corrections) 
\cite{Schonherr:2008av}, respectively. Throughout, all possible decays 
leading to the desired final state are considered, including all invisible 
$Z$- and hadronic $W$-, $Z$- and $\tau$-decay channels.

The distributions for the central values include hadronization 
\cite{Winter:2003tt} and an underlying event simulation \cite{Alekhin:2005dx}. 
The CT10 parton distributions \cite{Guzzi:2011sv} have been used throughout. 
Scales are set according to the \CKKW prescription \cite{Hoeche:2009rj,
  Hoeche:2012yf} and the uncertainties are evaluated as follows
\begin{itemize}
  \item To determine the renormalization scale the event is clustered using 
	the inverse of the parton shower, including electroweak splitting 
	functions as introduced in \cite{Denner:2006xxx,Krauss:2014yaa}, 
	until a $2\to 2$ core configuration is reached. The renormalization 
	scale $\mu_R$ is then defined through 
	\begin{equation}
	  \alpha_s^{k+n}(\mu_R)
	  \;=\;\alpha_s^k(\mu_\text{core})\prod\limits_{i=1}^n \alpha_s(t_i)\,,
	\end{equation}
        wherein $k$ is the QCD order of the such determined core process at 
        tree level, i.e.\ $k=0$ for $q\bar{q}^{(\prime)}\to VV^{(\prime)}$ or 
        $q\bar{q}^{(\prime)}\to VH$, $k=1$ for $q\bar{q}^{(\prime)}\to Vg$ or 
        $gq\to Vq^{(\prime)}$, and $k=2$ for pure QCD core processes. $n$ is the 
        final state clustered jet multiplicity and the $t_i$ their respective 
        reconstructed emission scales. As core scale we choose 
        $\mu_\text{core}=\hat{s}$ for $k=0$, $\mu_\text{core}=\tfrac{1}{2}\,
        m_\perp(V)$ for $k=1$, and $\mu_\text{core}=\tfrac{1}{2}\,p_\perp$ 
        for $k=2$. For $n=k=0$ we set $\mu_R=\mu_\text{core}$. The 
        factorization scale is set to $\mu_F=\mu_\text{core}$ on the core 
        configuration. The thus determined $\mu_R$ and $\mu_F$ are then varied 
        by a factor of 2.
  \item The resummation scale $\mu_Q$, also defined in~\cite{Hoeche:2012yf}, 
        is set equal to the factorization scale. It is varied by a factor of 
        $\sqrt{2}$, cf. \cite{Hoeche:2012fm,Hoeche:2014lxa}.
  \item $Q_{\rm cut}$ is the merging scale. Three values are chosen for this 
	scale, 15 GeV, 30 GeV and 60 GeV.
\end{itemize}
The uncertainties in all figures are shown as two bands, one for the combined
background and one for the combined signal, accumulated through their 
respective contributing processes only. They have been evaluated at the 
parton level. The full perturbative uncertainty for each process is obtained 
as the quadratic sum of the envelopes provided by the variation of the 
perturbative scales, $\mu_R$, $\mu_F$, and $\mu_Q$, and the merging scale 
$Q_{\rm cut}$. As non-perturbative uncertainties were found to be very small, 
these parton level uncertainties are directly applicable to the hadron level 
results. 
The electroweak input parameters for this simulation are $\alpha=1/128.802$,
$m_W=80.419~\mr{GeV}$, $m_Z=91.188~\mr{GeV}$ and $m_H=125~\mr{GeV}$.

\subsection{Results with \texorpdfstring{\protect{\MEPSatNLO}}{MEPS@NLO}}
\label{sec:res-pl}

\begin{figure}[t]
  \begin{center}
    \setlength{\unitlength}{1cm}
    \lineskip-1.7pt
    \includegraphics[width=16cm]{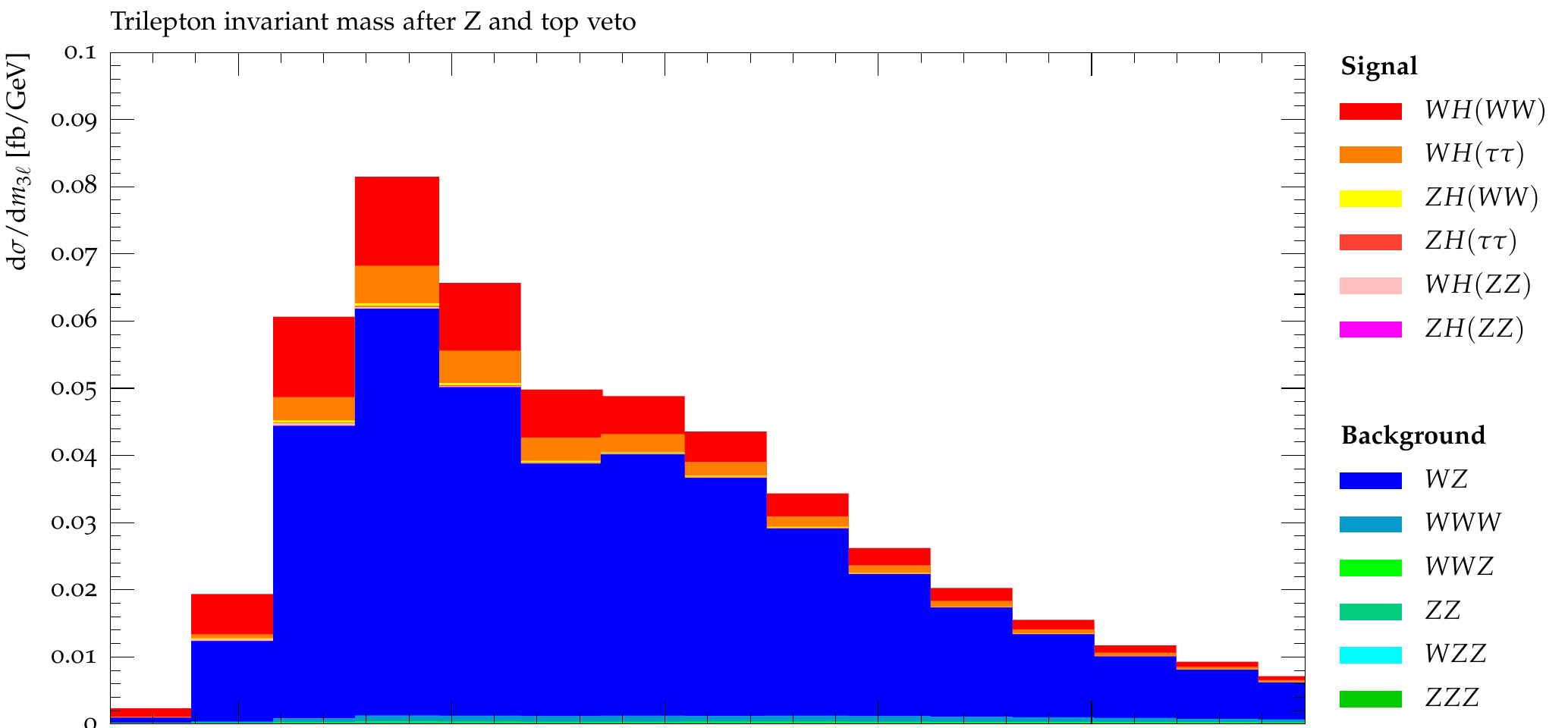}\\
    \includegraphics[width=16cm]{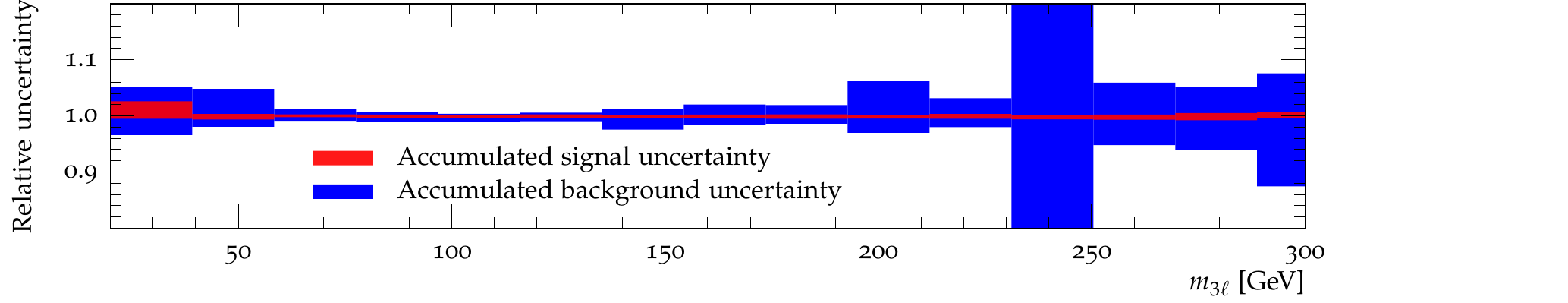}
    \begin{picture}(15,0)
      \put(6.3,6.4){\includegraphics[width=6.3cm]{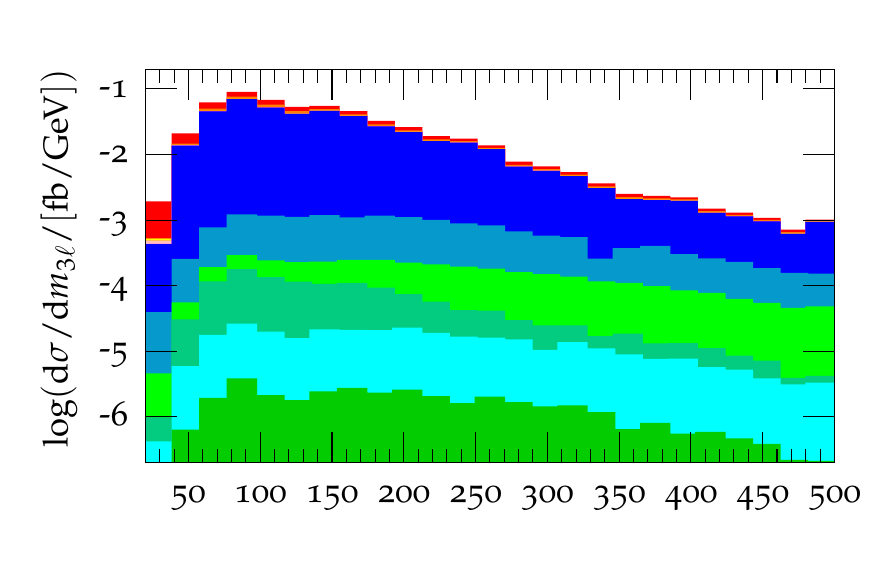}}
    \end{picture}
    \vspace*{-3mm}
    \caption{
	      The trilepton invariant mass after \protect\CMS cuts. All 
	      contributing processes, grouped as whether considered 
	      signal or background, are added incoherently, ordered by 
	      relative contribution. The inset displays the same information 
	      on a logarithmic scale to better quantify the contributions 
	      of the rarer processes. Below the main plot the accumulated 
	      relative uncertainties originating from the respective signal 
	      and background processes to the total expected cross section 
	      are detailed.
	      \label{fig:cms_mass}
	    }
  \end{center}
\end{figure}

This section presents selected observables defined on the event samples 
prepared with the analyses described in Sec.\ \ref{sec:res-ana} applied to the 
calculations of Sec.\ \ref{sec:res-mc}.  All observables considered below 
show a clear signal over background excess.  They focus on the leptons from the
hard process after the $Z$ and jet veto.  The $Z$ veto is very important in 
these analyses, as without it the $W^\pm Z$ process is very dominant over both 
the signal and the background, while without the jet veto top associated 
vector boson production would bury the Higgs processes.

The first observable we consider is the trilepton invariant mass of events 
in the \CMS-inspired analysis in Fig.~\ref{fig:cms_mass}.  
After the veto on the $Z$ boson and final state $b$-jets, the invariant mass
distribution of the 3 leptons can be used to distinguish the signal from the
background as a visible 30\% excess is seen in the peak region, far surmounting 
the background uncertainties displayed in the lower panel.  Very similar 
findings are made when looking at events in the \ATLAS-inspired analysis. 
Although the main signal process $W^\pm H(W^+W^-)$ forms the majority of the 
excess, the contribution from $W^\pm H(\tau\tau)$ is non-negligible, albeit of 
a slightly different shape.  Regarding the background processes, the tri-boson 
processes have a significantly harder $m_{3\ell}$ spectrum, raising their 
relative contribution in the high-mass region, as can be seen in the 
logarithmically plotted inlay.  

\begin{figure}[t!]
  \begin{center}
    \setlength{\unitlength}{1cm}
    \lineskip-1.7pt
    \includegraphics[width=16cm]{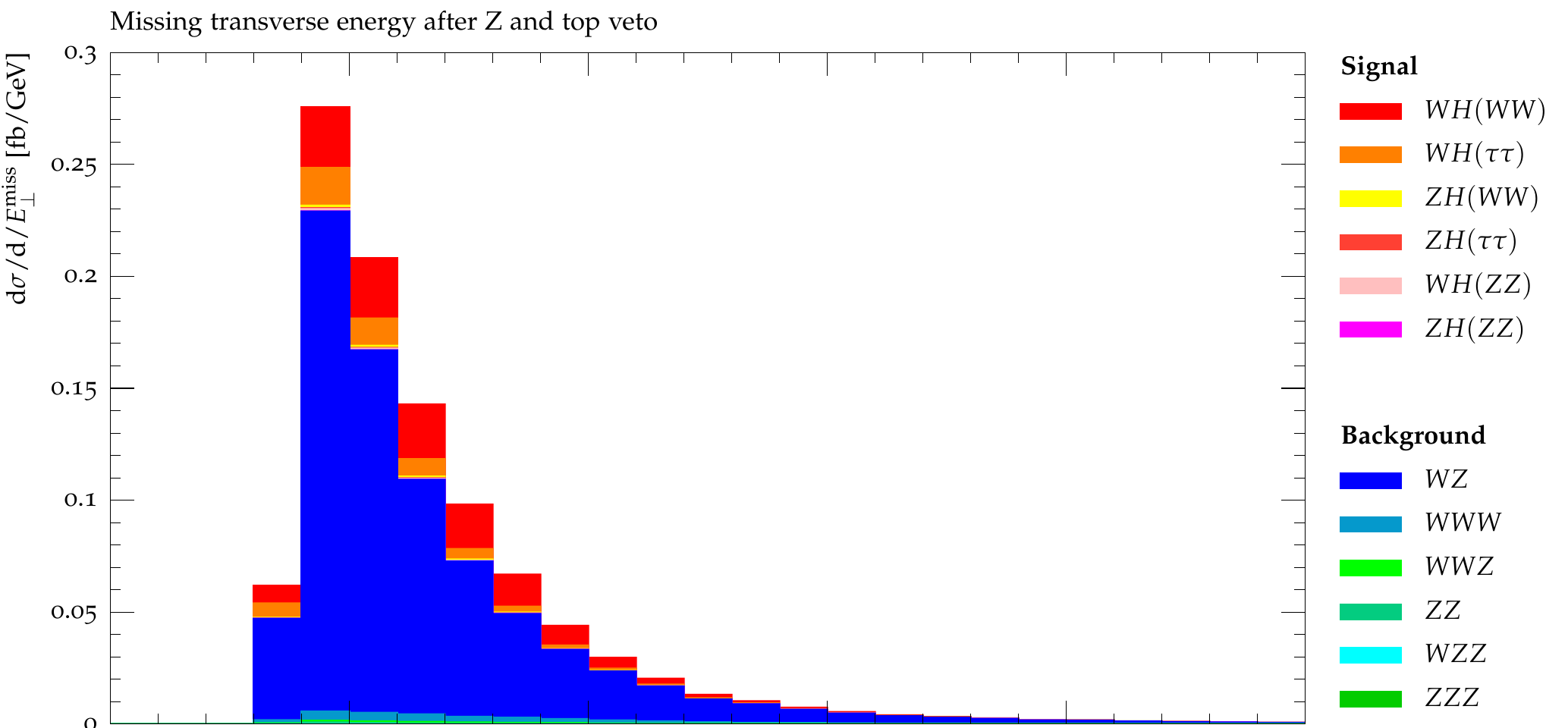}\\
    \includegraphics[width=16cm]{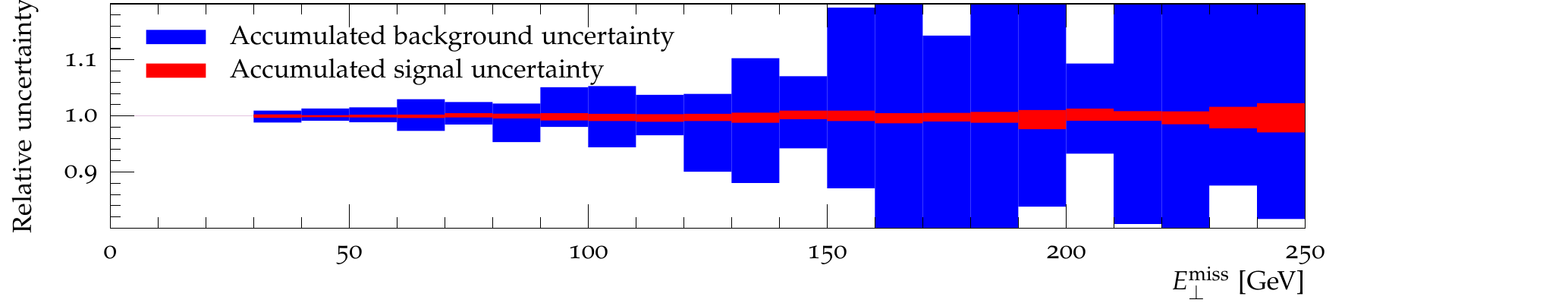}
    \begin{picture}(15,0)
      \put(6.3,6.4){\includegraphics[width=6.3cm]{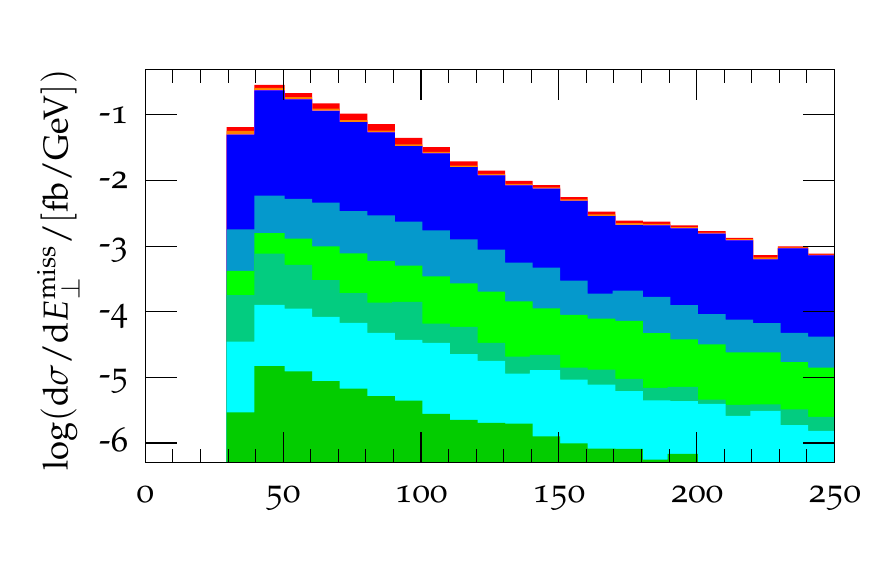}}
    \end{picture}
    \vspace*{-3mm}
    \caption{
	      The missing transverse energy spectrum after \protect\CMS cuts. 
	      For details, see Fig.\ \ref{fig:cms_mass}.
	      \label{fig:cms_etmiss}
	    }
  \end{center}
\end{figure}

A somewhat complementary observable is the missing energy distribution,
exhibited in Fig.\ \ref{fig:cms_etmiss}, again effected on the event 
selection of the \CMS-inspired analysis.  The findings indeed display a 
similar behavior to the trilepton invariant mass distribution of 
Fig.\ \ref{fig:cms_mass}, in that the signal is clearly visible above the 
background for $E_\perp^\text{miss}\lesssim\unit[100]{GeV}$.  Again, 
the dominant and subdominant signal processes, $W^\pm H(W^+W^-)$ and 
$W^\pm H(\tau\tau)$, exhibit somewhat different shapes, with 
$W^\pm H(\tau\tau)$ possessing less missing transverse momentum. In 
both observables, the $W^\pm Z$ background is the most dominant background. 
However, here the di-boson and tri-boson background have a very similar 
behavior at large $E_\perp^\text{miss}$.

\begin{figure}[p]
  \begin{center}
    \setlength{\unitlength}{1cm}
    \lineskip-1.7pt
    \includegraphics[width=16cm]{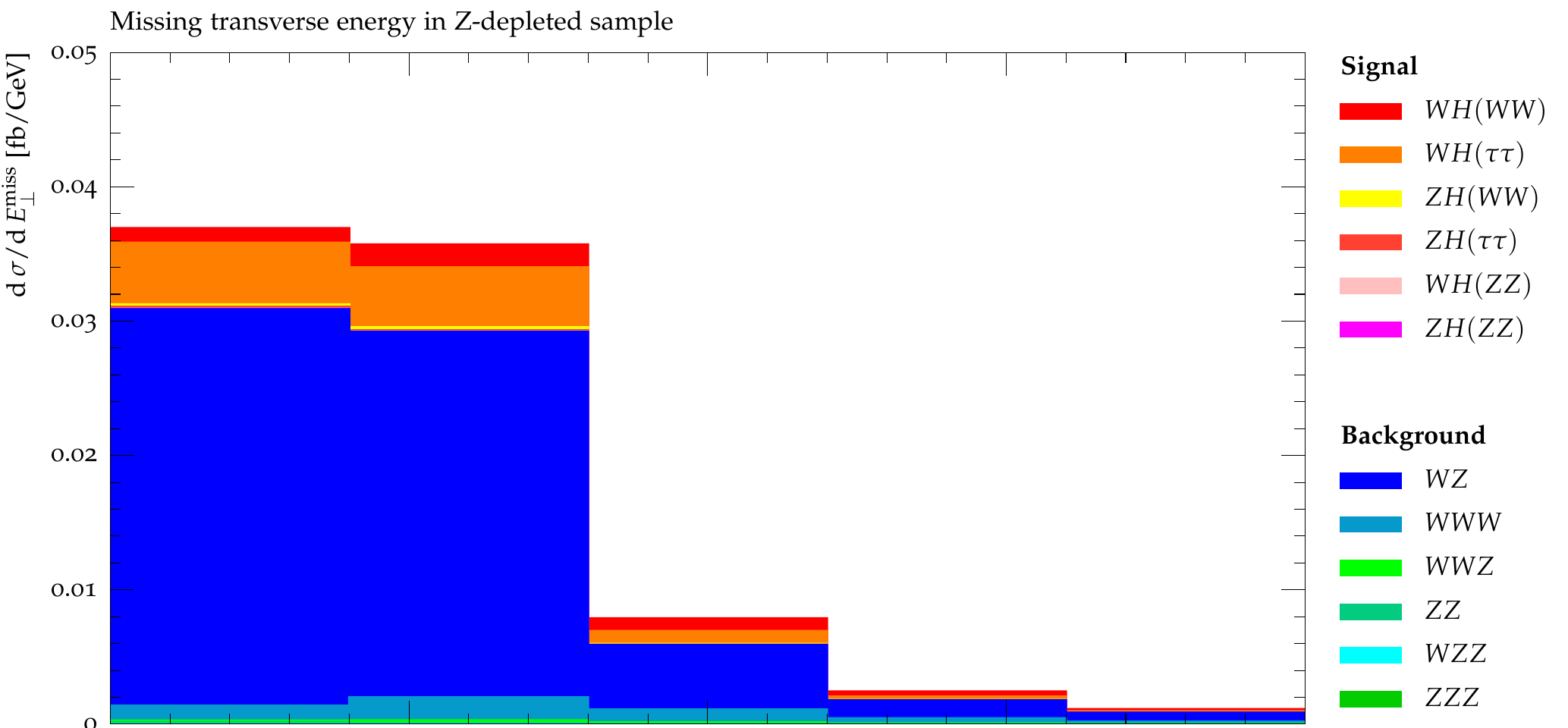}\\
    \includegraphics[width=16cm]{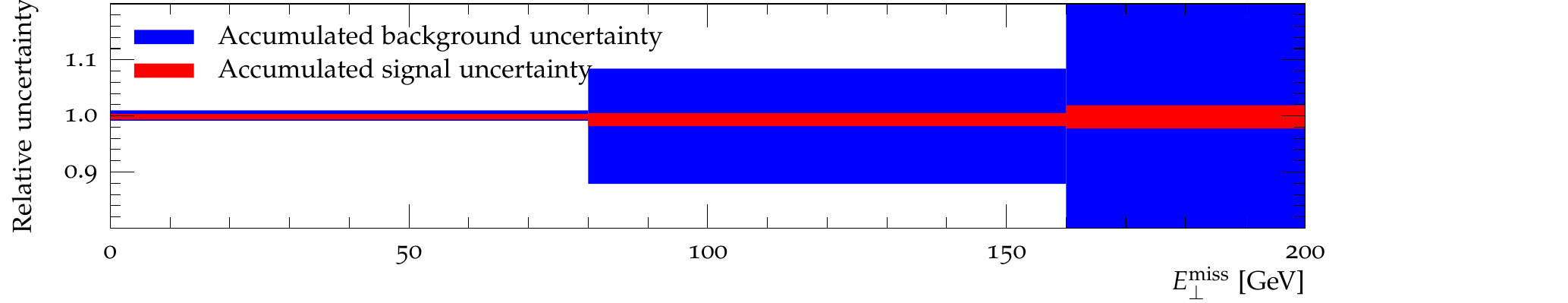}
    \begin{picture}(15,0)
      \put(6.3,6.4){\includegraphics[width=6.3cm]{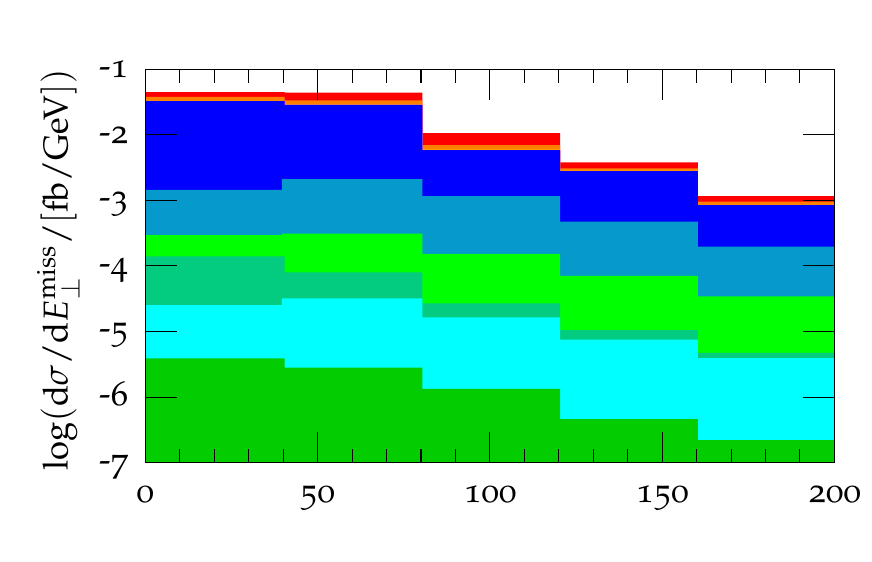}}
    \end{picture}
    \vspace*{-4mm}
    \caption{
	      The missing transverse energy spectrum after \protect\ATLAS cuts. 
	      For details, see Fig.\ \ref{fig:cms_mass}.
	      \label{fig:atlas_etmiss}
	    }
     \end{center}
\end{figure}

\begin{figure}[p]
  \begin{center}
    \setlength{\unitlength}{1cm}
    \lineskip-1.7pt
    \includegraphics[width=16cm]{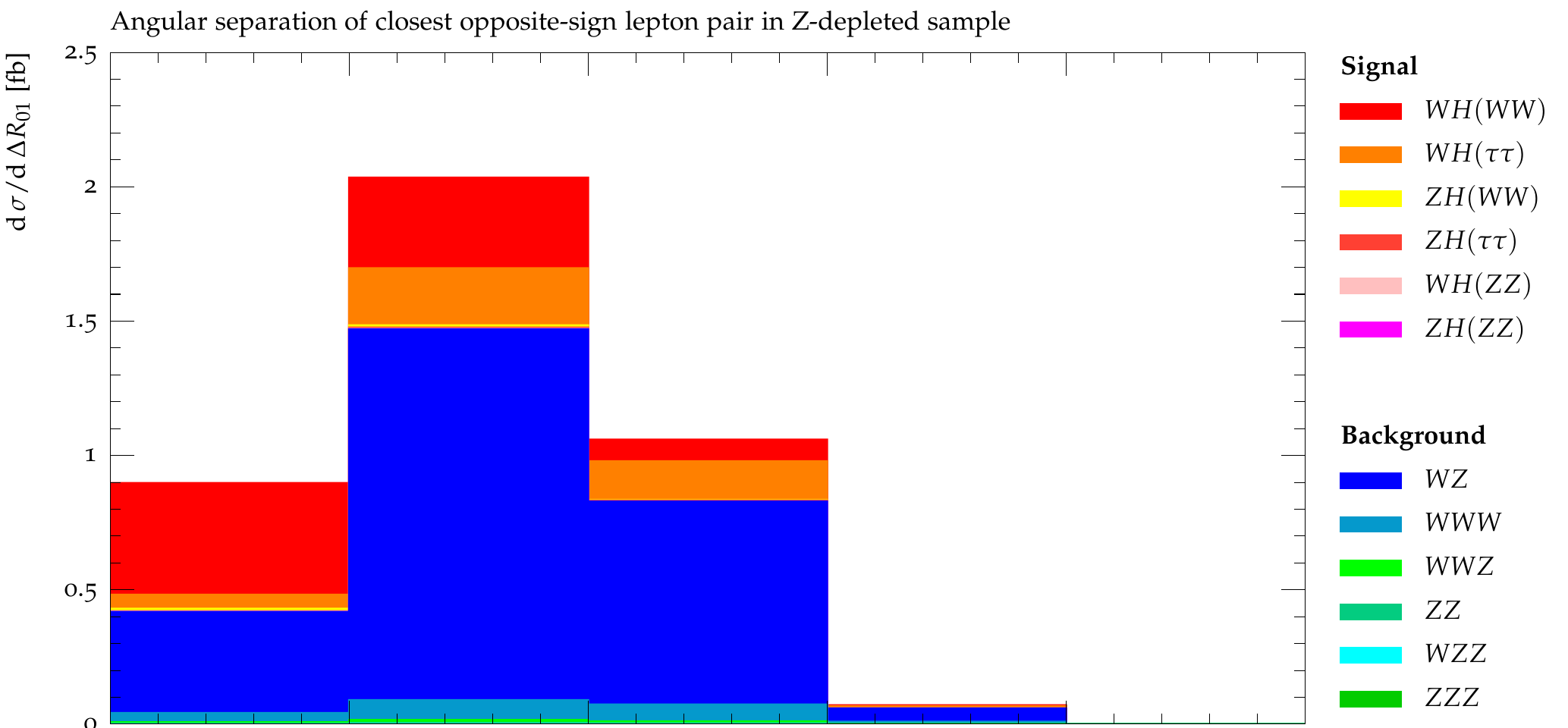}\\
    \includegraphics[width=16cm]{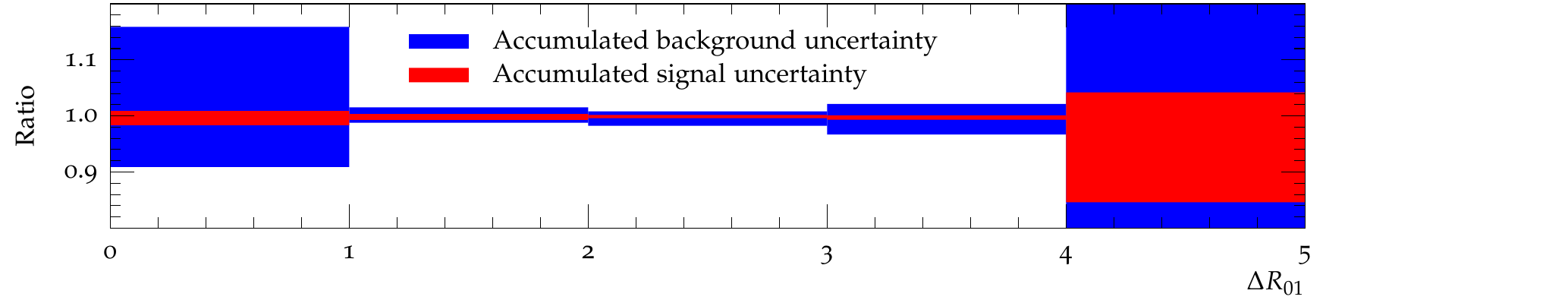}
    \begin{picture}(15,0)
      \put(6.3,6.4){\includegraphics[width=6.3cm]{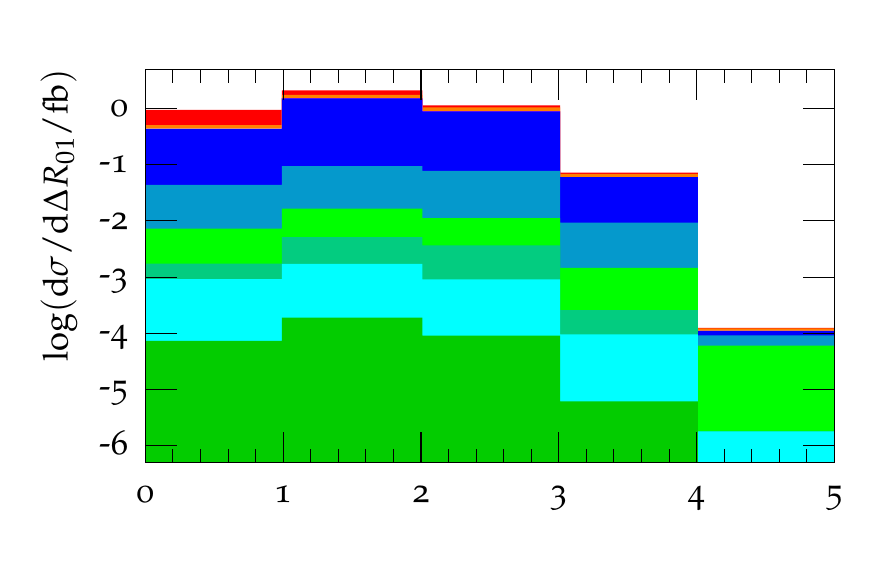}}
    \end{picture}
    \vspace*{-4mm}
    \caption{
	      The angular separation of the closest pair of oppositely 
	      charged leptons in the case that no SFOS pair of leptons 
	      is found in the event after \protect\ATLAS cuts. 
	      For details, see Fig.\ \ref{fig:cms_mass}.
	      \label{fig:atlas_dr}
	    }
  \end{center}
\end{figure}

The relatively small excess in the $E_\perp^\text{miss}$ spectrum in the 
\CMS-inspired event selection is enhanced in the \ATLAS-inspired event 
selection with its stronger $Z$ veto, implemented through a complete rejection 
on SFOS lepton pairs.  Here, the $E_\perp^\text{miss}$ distribution 
shows an excess of the signal over the background of up to 50\%.  This is 
displayed in Fig.\ \ref{fig:atlas_etmiss}.  In contrast to the case of a $Z$ 
veto through a mass window as in \CMS, where the distribution especially
for $W^\pm Z$ falls off smoothly, here the $Z$ veto introduces a visible kink,
while the signal remains unaffected.  This of course could be further used
to reduce the $W^\pm Z$ background by utilizing this different impact on the 
respective shapes.

The angular separations between pairs of leptons are interesting observables 
for this process.  Fig.\ \ref{fig:atlas_dr} shows the distance $\Delta R$ 
between the closer of the two pairs of oppositely signed leptons, following 
the \ATLAS-inspired event selection. These leptons do not have the same flavor, 
as this observable isolates the leptons that are most likely to be products of 
the Higgs boson decay to $W^+ W^-$ or $\tau$ pairs.  This effect in particular 
on the $WW$ channel stems from the spin correlations in the decay of the Higgs 
boson, as already discussed in~\cite{TheATLAScollaboration:2013hia}. As a 
result, this observable also has good discriminating power between signal and 
background, providing a clear excess in the region $\Delta R<3$. It also, 
better than the other observables considered, separates the two main signal 
processes. While $W^\pm H(W^+W^-)$ constitutes approximately $\unit[80]{\%}$ 
of the Higgs signal below $\Delta R=1$, $W^\pm H(\tau\tau)$ contributes 
roughly $\unit[60]{\%}$ in the region $2<\Delta R<3$.  There, however, the 
signal excess over the background has fallen from a factor of two 
to approximately $\unit[35]{\%}$ of the background expectation.

The different uncertainties have been investigated individually for all 
processes to check for their dominant source.  In nearly all bins of all
observables considered here, the uncertainties are driven by the
renormalization and factorization scale variation with a typical effect on the
few--\-percent level up to about 10\% for the tri-boson processes.  In
regions dominated by jet activity of course the \MENLOPS samples, being at
leading order accuracy only, exhibit a stronger dependence than those processes
simulated with \MEPSatNLO.  In addition, it is worth stressing that effects
due to hadronization and the underlying event are practically irrelevant for
the uncertainties in the simulation of the processes for the observables 
considered here.  Their main effect is on the isolation efficiency of the 
leptons. Although the non-perturbative corrections have a clear impact on the 
shape of trilepton invariant mass of Fig.\ \ref{fig:cms_mass}, as the 
isolation is $p_\perp$-dependent, their uncertainties are barely noticeable. 
On the contrary, the missing transverse energies of Figs.\ \ref{fig:cms_etmiss} 
and \ref{fig:atlas_etmiss} and the angular separation of Fig.\ 
\ref{fig:atlas_dr} receive merely a change of the overall rate from effecting 
non-perturbative corrections. Again, their uncertainties are negligible.

\section{Conclusions}
\label{Sec:Conclusions}
In this publication NLO QCD accurate predictions for multiple weak 
boson production at the LHC were presented, and their application to Higgs boson
searches based in trilepton final states has been highlighted. 
The $W^\pm H$ and $ZH$ Higgsstrahlung signals as well as the main 
backgrounds, $W^\pm Z$ and $W^\pm W^+W^-$ production, have been simulated at 
NLO including up to one extra jet in the \MEPSatNLO multi-jet 
merging framework.
The simulation of the $W^\pm W^+W^-$ background represents a nontrivial
application of multi-jet merging at NLO 
and plays and important role for all Higgs and new-physics searches based on 
trilepton final states and jet vetoes.
Also various other diboson and tribosons background processes have 
beed computed at NLO QCD, including
matching to the parton shower and an improved description of 
extra jet radiation, based on the MENLOPS technique.
We confirm, at NLO, that the relevant backgrounds to $W^\pm H$ and $ZH$ production
are given by diboson and triboson production processes, if jet vetoes can be
applied.  We also show that the residual perturbative uncertainties in large
fractions of the relevant phase space are of the order of 10\% or even below.
This will offer excellent opportunities for Higgs boson precision studies at the
forthcoming LHC runs.

\section*{Acknowledgments}
We are grateful to A.~Denner, S.~Dittmaier and L.~Hofer for providing us 
with the one-loop tensor-integral library \Collier.

This work has been supported in part by the European Commission through 
the ``LHCPhenoNet'' Initial Training Network PITN-GA-2010-264564, through
the ``MCnet'' Initial Training Network PITN-GA-2012-315877, and through
the ``HiggsTools'' Initial Training Network PITN-GA-2012-316704.
SH was supported by the U.S.\ Department of Energy under Contract 
No.\ DE--AC02--76SF00515. SP was supported by the SNSF.
This research used resources of the 
National Energy Research Scientific Computing Center, which is supported 
by the Office of Science of the U.S. Department of Energy under Contract 
No. DE--AC02--05CH11231.

\bibliographystyle{bib/amsunsrt_modrc}
\bibliography{bib/journal}
\end{document}